\documentstyle[sprocl,epsf,epsfig,wrapfig]{article}

\def\Journal#1#2#3#4{{#1} {\bf #2}, #3 (#4)}

\def\PLB{{\em Phys. Lett.}  B}
\def\PRL{\em Phys. Rev. Lett.}
\def\PRD{{\em Phys. Rev.} D}

\def\be{\begin{equation}}
\def\ee{\end{equation}}
\def\bea{\begin{eqnarray}}
\def\eea{\end{eqnarray}}

\begin{document}

\title{Plans for Spin Physics at RHIC}

\author{L.C. Bland}

\address{Indiana University Cyclotron Facility, 2401 Sampson 
Road, Bloomington,\\ IN 47408, USA\\E--mail: bland@iucf.indiana.edu} 


\maketitle
\abstracts{ Polarized proton collisions will be studied at RHIC 
up to a total center of mass energy of 500 GeV, starting in 2002.  An
overview of the RHIC-spin program, and the critical
components of the PHENIX and STAR detectors for spin experiments, is
presented.  The premier experiment within the RHIC-spin program is the
determination of the fraction of the proton's spin carried by gluons.
A detailed analysis of how accurately this fraction can be determined
by the STAR experiment at RHIC is presented.}


\bigskip
\section{Introduction}

The purpose of this workshop is to establish the 
primary physics goals of a polarized 
electron-polarized ion collider (EPIC) that 
would become operational, at the earliest, in 2005.  To
predict what the most interesting questions will 
be {\it in the future}, it is necessary to understand 
what questions are being addressed in the present, 
and to establish the expected quality of the 
answers.  With that perspective, I will describe 
plans for an experimental program at Brookhaven 
National Laboratory that will study the high-energy collisions 
of polarized protons, 
accelerated in the Relativistic Heavy Ion Collider (RHIC).  
There is great expectation within the 
community that the RHIC-spin program will provide 
us with important information about the spin 
structure of the proton.  Arguably, the primary goal 
of the RHIC-spin experiments is to determine 
the fraction of the proton's spin carried by gluons (the integral
$\Delta G$, defined below).  
Establishing the degree of polarization of the 
glue within the proton is the {\it next essential step} 
in understanding the spin structure of the 
proton.  Under the assumption that experiments establishing 
the gluon polarization are the most 
important, most of this talk will be devoted to that subject.

\section{An overview of the RHIC spin program}
Before delving into the details about measuring 
the polarization of the proton's glue, it is important 
to give a broader overview of the physics of 
RHIC spin.  To date, the highest energy achieved in a 
synchrotron for a polarized proton beam is 24.6 GeV 
in the AGS \cite{BAI98}.  The RHIC rings will provide polarized 
proton beams up to a maximum energy of $\sim$250 GeV, 
an order of magnitude higher in energy.  Even more 
impressive is that the RHIC-spin program will 
provide total center of mass 
energy ($\sqrt{s}$) for collisions between polarized 
protons up to 500 GeV.  Furthermore, the expected 
luminosity for ${\vec p}+{\vec p}$ collisions at 
RHIC ($2 \times 10^{32}$cm$^{-32}$s$^{-1}$ at $\sqrt{s}$ = 500 GeV), 
will enable a systematic study of large transverse 
momentum ($p_T$) processes, where perturbative QCD has 
been successfully applied to explain much of the data 
obtained at {\it unpolarized} $pp$ and $p \overline{p}$ 
colliders.  The addition of polarization to high-$p_T$ 
collisions between protons at 
very high energies will provide an important test of QCD.

With the addition of polarization to colliding 
proton beams, {\it spin asymmetries} ($A$):
$$PA={N_+ - N_- \over N_+ + N_-}, \eqno(1)$$
are new observables that can be measured.  In Eqn.\ 1, 
the beam polarization(s) are represented by 
$P$ and are expected to be 70\% for the RHIC-spin program.  
Asymmetry measurements are made by determining how the yield 
for some process varies with the polarization state ($N_\pm$) of the beam(s)
and by measuring the polarization of the beam(s).  Eqn.\ 1 assumes
equal integrated luminosities for the two polarization states of the beam(s).
For the RHIC-spin program, it is expected 
that transverse single-spin asymmetries ($A_T$) will 
approach zero for sufficiently large $p_T$; non-zero 
values for $A_T$ are expected only from higher-twist 
contributions.  Parity-violating longitudinal single-spin asymmetries 
($A_L$) are expected to be quite large in the production of 
{\it real} weak bosons ($W^\pm,Z^0$) \cite{BS93} or at sufficiently 
high $p_T$, where {\it virtual} weak bosons contribute 
significantly \cite{TV95} to the force between the interacting partons.

Double-spin asymmetries involve differences of yields for a 
process when the initial state protons have the same and 
opposite direction polarizations.  The (parity-allowed)
longitudinal asymmetries ($A_{LL}$)
are expected to teach us about helicity asymmetry structure 
functions, similar to those probed in polarized deep 
inelastic scattering.
Transverse double-spin asymmetries ($A_{TT}$) may enable 
measurement of the transversity distributions \cite{JA97}, related to 
the transverse polarization of the nucleon's constituents.

If QCD survives the onslaught of these new high-energy 
and high-$p_T$ polarization observables,
then polarized proton collisions can 
be used as a tool to better understand the spin structure of 
the proton.  As described in detail below, $A_{LL}$ is very
sensitive to the gluon polarization for several different processes.
In addition, parity-violating $A_L$ measurements for the 
production of $W^\pm$ and $Z^0$ can be directly 
related to the polarization of valence and sea quarks 
within the proton \cite{BS93}.  Since the valence quark 
polarizations are well determined in polarized deep 
inelastic scattering, direct tests of the Standard 
Model are possible.  By changing the kinematical conditions, 
${\vec p}+p\rightarrow W^\pm+X$
can be used to selectively probe antiquark polarizations, 
thereby providing new information about 
the proton's spin structure.  Polarization observables for the 
highest $p_T$ processes at RHIC may also provide limits on 
either the existence of new vector bosons or possible quark 
substructures \cite{TV95}, competitive with planned 
measurements at the Tevatron.  

\begin{wrapfigure}{r}{6cm}
\epsfig{figure=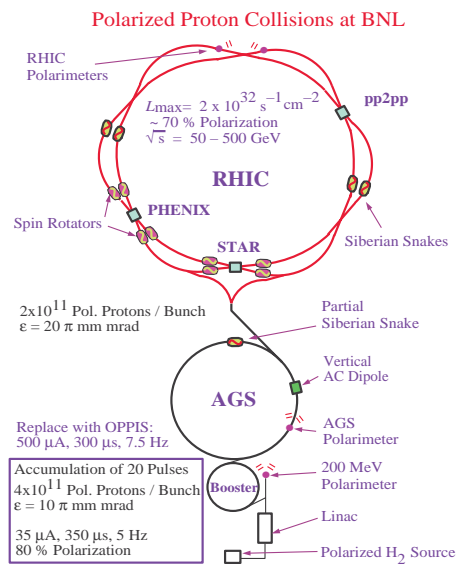,width=6cm}
\caption{Schematic layout of the RHIC accelerator complex, indicating
the critical elements needed for the acceleration of polarized ions.}
\end{wrapfigure}

\section{Preparing RHIC for polarized proton studies}
The RHIC spin program will present the opportunity 
to study observables in polarized $pp$ collisions 
at $\sqrt{s}$ values that are more than {\it a factor of 25
larger than previously available}~\cite{Ad91}.  This frontier 
of high-energy spin physics is made technically feasible by significant 
advances in the handling of polarized beams in 
synchrotrons and storage rings.  The new technique 
employs so-called `Siberian snakes', a concept 
introduced in Novosibirsk \cite{DK78} and first experimentally 
verified at the IUCF Cooler ring \cite{Kr89}.  The Siberian snake 
is a means of overcoming the large number 
of depolarizing resonances encountered when attempting to 
accelerate polarized proton (or other 
light ion) beams to high energies in a circular 
accelerator.  For RHIC, helical dipole magnets,
funded by the RIKEN institute in Japan, will be used to
reverse the direction of unwanted transverse polarization 
components during alternate turns around each ring.
In the absence of such spin 
manipulations, it is possible to destroy the 
beam polarization by small perturbations that act in a 
single turn, with their effects adding 
coherently when the ring is operated near a depolarizing 
resonance.

The motivation to build RHIC and its two major 
detectors, PHENIX and STAR, originated from 
the goal of discovering and studying a new state of 
matter, the {\it quark-gluon plasma}, that is 
thought to have existed at the earliest moments 
after the Big Bang.   It is expected that the quark-gluon 
plasma should be formed in the ultrarelativistic 
collisions between heavy-ion beams at 
RHIC.  Although the major detectors were not designed for 
the study of ${\vec p}+{\vec p}$ 
collisions up to total center of mass energy equal to 
500 GeV, many of their subsystems have 
been subsequently adapted to this task.  As a result, 
PHENIX and STAR bring complementary strengths to the 
RHIC-spin program.  

The PHENIX detector has many relatively small acceptance 
detectors that provide fine granularity 
and precise particle identification.  Of greatest relevance 
to the spin program are the photon arms that are
centered at midrapidity ($|\eta|\leq 0.35$), and the 
muon arms, spanning $1.0\leq |\eta| \leq 2.4$.  
The fine granularity of the Pb-glass and Pb+scintillating 
fiber elements of the PHENIX electromagnetic calorimeter will 
provide the capability to separately identify single photons and 
di-photon pairs, produced in the decay of neutral mesons.  
These detectors will be employed for inclusive photon 
production in the RHIC-spin program.  As well, they will be used to
study high-$p_T ~ \pi^0$ production.  The muon arms provide 
excellent particle identification, and will be used for the 
study of $W^\pm$ production, by observing the $\mu^\pm$
daughters of the $W$.  In addition, the detection of $\mu^+\mu^-$ 
pairs with the PHENIX muon arms will enable a program of 
measuring polarization observables associated with vector 
meson production and the Drell-Yan process.

The STAR detector is intended to provide a more global view 
of a heavy-ion collision.  The heart of STAR is a 0.5 T solenoidal 
magnetic field and a time projection chamber (TPC), capable of tracking all 
of the charged particles produced in a central Au-Au collision 
in the nominal pseudorapidity interval, $|\eta|\leq2$, with full 
azimuthal coverage.  Multiple layers of silicon detectors around the 
interaction point will be used for reconstructing primary event 
vertices, and secondary vertices from strange particles.  Of 
greatest relevance for the RHIC-spin program are the barrel and endcap 
electromagnetic calorimeters (EMC).  Construction of the barrel 
EMC is underway and should be completed by 2002.  
A proposal \cite{CDR99} for the endcap EMC is awaiting final decision 
concerning funding.  If approved, the construction timetable for
that detector will be comparable to the barrel EMC project. 
In comparison to PHENIX, 
the STAR EMC has substantially coarser granularity, but provides full 
azimuthal coverage for the pseudorapidity interval, 
$-1\leq \eta \leq 2$, with a small gap near $\eta = +1$ for TPC and EMC 
services.  The large acceptance of STAR makes it ideal 
for jet reconstruction.  As described below, detection of 
photon + jet coincidences provides STAR with unique capabilities 
in determining $\Delta G$.  In addition, measurements of di-jet production at 
STAR will provide an important cross check of 
the gluon polarization measured in direct photon production.  
Finally, the STAR EMC will also enable the study of $W^\pm$ 
production, by observing the $e^\pm$ daughters of the $W$.

As of July 1999, the status of the RHIC accelerator is as follows.  
Commissioning of Au beam in the RHIC rings began this summer.
The initial studies of Au-Au collisions will commence in
November 1999.  The first helical dipole magnets have been successfully
produced, and the initial commissioning of a polarized proton beam is
scheduled for 2000.  Initial studies of $\vec{p}+\vec{p}$ collisions
at low luminosity are planned for 2001.  
It is expected that full luminosity ${\vec
p}+{\vec p}$ collisions will begin in 2002.  An agreement
between RIKEN and BNL will allow the RHIC-spin program to run for 10
out of the 37 weeks of annual RHIC operations.  It is projected that
at full luminosity, a 10-week run at $\sqrt{s}$=200(500) GeV will
result in an integrated luminosity of 320(800) pb$^{-1}$.

\section{Methods of determining $\Delta G$}
After this brief overview of the spin-physics program and the tools to
carry it out, I'll focus the rest of this writeup on the 
determination of the fraction of
the proton's spin carried by gluons, which is expected to be the most
important result forthcoming from the RHIC-spin program.
That fraction is equal to twice the integral of the 
gluon helicity asymmetry distribution ($\Delta G(x,Q^2)$):
$$\Delta G(Q^2) = \int_0^1 \Delta G(x,Q^2) dx = 
\int_0^1 [ G^+(x,Q^2)-G^-(x,Q^2)] dx, \eqno(2)$$
and is a function of 
the scale, $Q^2$.  For convenience, I will
subsequently suppress the dependence on $Q^2$.  The asymmetry 
is given by the difference in probability of finding a gluon 
with its polarization parallel ($G^+$) versus 
antiparallel ($G^-$) to the proton's longitudinal polarization.  
The unpolarized gluon distribution function is given 
by $G(x)=G^+(x) + G^-$(x).  The gluon polarization is given by the
ratio, $\Delta G(x) / G(x)$.  Similar definitions 
of unpolarized parton distribution functions 
and helicity asymmetry distributions exist for quarks 
($q$) and antiquarks ($\overline{q}$).

\begin{figure}
\epsfxsize=11.5cm \epsfbox{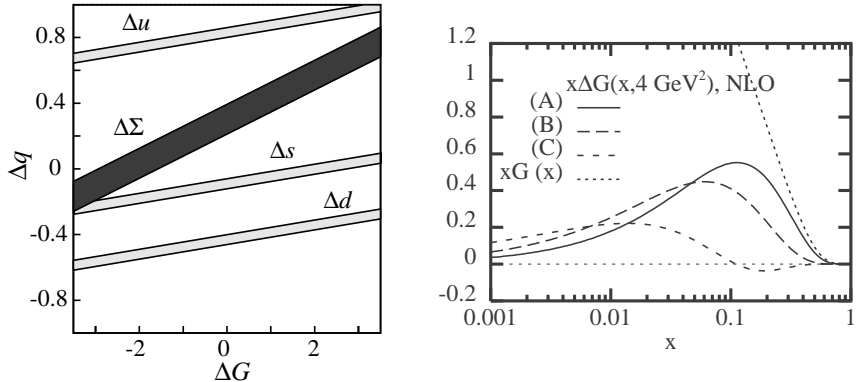}
\caption{(Left) Correlation between the quark and gluon contributions
to the proton's longitudinal polarization.  (Right) Three models$^{10}$ 
of the gluon helicity asymmetry distribution.}
\end{figure}

The importance of $\Delta G$ is twofold.  First, the proton's 
spin can be decomposed as follows:
$$S_z ~=~ {1\over 2} ~=~ {1\over 2}\Delta \Sigma ~+~ \Delta G ~+~ L_z^q
~+~ L_z^G . \eqno(3) $$
Here, $L_z^{q(G)}$ gives the orbital angular momentum 
contributions of quarks (gluons) to the 
proton's spin.  The contribution quarks make to 
the proton's spin ($\Delta \Sigma$) is determined from quark 
helicity asymmetry functions, $\Delta q_i(x)$, summed over the 
quark flavors ($n_f$) consistent with $Q^2$, by: 
$$ \Delta \Sigma = \sum_{i=1}^{n_f} \Delta q_i, {\rm where~~} 
\Delta q_i=
\int_0^1 [q_i^+(x)+\overline{q}_i^+(x)
-q_i^-(x)-\overline{q}_i^-(x)]dx. \eqno(4)$$
Hence, if $\Delta \Sigma$ is known, then a 
determination of $\Delta G$ establishes the 
contribution of partonic orbital motion to the 
proton's spin.  

The second reason for the importance 
of $\Delta G$ is that, due to the axial anomaly 
of QCD, the quantity $\Delta \Sigma$ {\it cannot be 
determined independently} of $\Delta G$ in polarized deep inelastic
scattering (PDIS).  
This is graphically illustrated in Fig. 2, 
showing the results from the global 
analysis of polarized deep-inelastic scattering (PDIS) 
made by the SMC group \cite{SMC97}.
The analysis of the limited information on scaling 
violations in PDIS \cite{GS96} has produced
only crude constraints on $\Delta G(x)$ and its integral.

The question then is, how can $\Delta G$ be determined?  There are
several possible methods being pursued:
\begin{itemize}
\item Measurements of PDIS spanning a broad range of $x$ and $Q^2$
could be performed.  A determination of $\Delta G(x)$ would result
from analysis of the scaling violations from this more extensive data
set.  This method would require a
high-energy polarized $ep$ collider.  Possible plans for pursuing such
a program at HERA were discussed at this workshop \cite{dR99}.

\item Measurement of the leptoproduction of di-jet events is sensitive
to $\Delta G(x)$ through the photon-gluon fusion process.  A variation
of this method is to detect high-$p_T$ charged hadron pairs, assumed to
be the leading particles of jets, in high-energy
polarized-lepton/polarized-proton collisions.  This method is being
pursued by COMPASS \cite{BHK98}, and intriguing data has already been
obtained at much lower energy by HERMES \cite{Br99}.

\item Measurements of di-jet production, high-$p_T$ particle
or photon production in ${\vec p}+{\vec p}$ collisions are sensitive
to $\Delta G(x)$.  This is the primary focus of the RHIC-spin program.

\end{itemize}

Sensitivity to $\Delta G(x)$ in polarized proton collisions 
arises for partonic collisions 
involving gluons.  Di-jet production can be initiated 
either by $qq$, $qg$ or $gg$ collisions, with 
the latter two possibilities involving $\Delta G(x)$  
linearly or quadratically, respectively.  Just as for 
lepton-induced processes, 
detection of high-$p_T$ hadrons, assumed to be the particles of
final-state jets, is also sensitive to $\Delta G(x)$.  But, the
relationship between the initial-state partonic kinematics and the
hadron's $p_T$ is more complex because of the momentum sharing between
the multiple hadrons within the jets.

Direct photon production is, in principle, 
the cleanest probe of $\Delta G(x)$ for ${\vec p}+{\vec p}$ collisions, since 
in leading-order pQCD, observables are only linearly dependent 
on the gluon structure function for 
the dominant gluon Compton scattering process $gq\rightarrow \gamma q$.  For 
$pp\rightarrow \gamma X$, there is only a 
small physics background from $q\overline{q}\rightarrow \gamma g$.
Furthermore, when the photon is detected in coincidence with the 
away-side jet, the initial-state partonic 
kinematics can be reconstructed (see Sect. 5.3).  
Hence, direct photon production is a 
primary focus of the RHIC spin program.

\section{Plans for determining $\Delta G$ at STAR}
One of the most promising methods for determining 
$\Delta G$ is to study direct photon 
production in polarized proton collisions.  Both PHENIX 
and STAR will study this process for the 
RHIC-spin program.  Below, I give a detailed assessment, 
somewhat biased towards the performance of the STAR detector, 
of how accurately $\Delta G$ will be determined in polarized 
proton collisions at RHIC.

\subsection{General features of direct photon production in $pp$ collisions}
Within the framework of leading-order perturbative QCD, the 
dominant mechanism for producing a single photon with large 
transverse momentum in a $pp$ collision is gluon Compton scattering
($gq\rightarrow \gamma q$).  The competing partonic 
subprocess is $q\overline{q}$ annihilation ($q\overline{q}\rightarrow
\gamma g$).  
Given the predominance of gluons over antiquarks at small Bjorken 
$x$, gluon Compton scattering is ten times more likely to 
occur.  Fig. 3 shows the pQCD predictions for the angular 
dependence of the cross section and longitudinal spin correlation 
coefficient for these two processes in the partonic center-of-mass 
reference frame.  The polarized cross section ($\Delta {\hat \sigma} =
{\hat a}_{LL}{\hat \sigma}$) is strongly peaked 
at scattering angles where the photon is emitted in the direction 
of the incident quark.  This result is independent of the total 
energy in the partonic CM frame.  The best determination of the 
gluon polarization will be made when the final-state photon is
parallel to the initial-state quark;
at other angles, there is reduced sensitivity to the 
gluon polarization.

\begin{wrapfigure}{r}{4.5cm}
\epsfig{figure=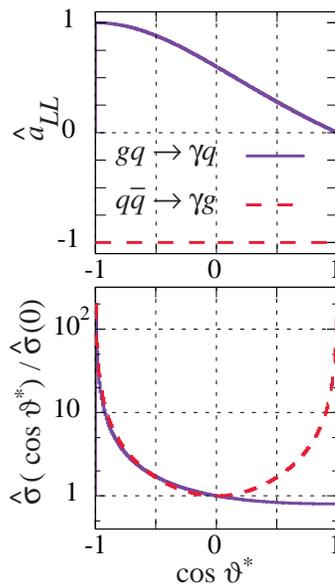,width=4.5cm}
\caption{Leading-order perturbative QCD predictions for the partonic
spin correlation coefficient and the relative differential cross
section for photon production processes.  For gluon Compton
scattering, the partonic center of mass scattering angle ($\theta^*$)
is defined relative to the incident gluon.}
\end{wrapfigure}
A second important criterion for optimizing the 
sensitivity to the gluon polarization is to require large 
polarizations for the initial-state quark.  Measurements \cite{Pe99} of 
polarized deep-inelastic scattering (DIS) have determined that the 
quark polarization increases linearly with log $x$,
for momentum fractions greater than 0.1.  At 
$x_q \approx 0.2$, the quark polarization is $\sim 0.3$, 
and it continues to rise as $x_q$ 
increases.  Furthermore, we have learned from 
{\it unpolarized} DIS, that the small $x$ region 
($x_g \leq 0.1$) is precisely where gluons 
predominately reside \cite{dR99}.  Hence, the optimum determination of 
the fraction of the proton's spin carried by gluons 
will be made by using large-$x$ quarks ($x_q \geq 
0.2$) as an {\it analyzer of the polarization} of small-$x$ gluons.  

These two criteria suggest that the greatest sensitivity to
$\Delta G(x)$ results from {\it asymmetric gluon Compton scattering} ($x_q >
x_g$), with the photon detected in the same direction that
the partonic CM moves in the collider reference frame.
As a consequence, both the photon and the 
hadronic jet from gluon Compton scattering events 
should be detected at large pseudorapidity, 
limited only by the need for large $p_T$ in the 
collision, generally assumed to be proportional to 
the kinematic scale, $Q$, relevant for the structure 
functions.  To limit the contributions from 
higher-twist processes, we will consider below 
direct photon production with $p_{T,\gamma} \geq 10$ GeV/c.

\subsection{Simulating polarized proton collisions}
Sophisticated event generators have been 
developed for high-energy (unpolarized) $pp$ and $p\overline{p}$
collisions to aid in the optimization of 
experiments.  To explore questions about the 
expected performance of the STAR detector we turn to a 
QCD event generator (PYTHIA 5.7) \cite{SJ94}, 
known to include many of the salient features of 
hard scattering processes. Hadronization of the 
recoiling quarks and gluons following a hard-scattering 
event is accounted for by the Lund string 
model, with parameters tuned to agree with fragmentation 
functions measured in $e^+e^-$ colliders.  The multiple 
soft-gluon emission, thought to be responsible for the introduction of 
transverse momentum to the initial-state partons, 
is accounted for in PYTHIA through the parton-shower 
model \cite{SJ85}, rather than by explicit evaluation 
of higher-order QCD processes.  

Spin effects are included for each event by 
evaluating the leading-order pQCD expressions for the 
process-specific spin-correlation coefficient \cite{BS89}, 
${\hat a}_{LL}^{proc}$, using the Mandelstam 
variables (${\hat s},{\hat t},{\hat u}$) for the 
partonic hard scattering, as given by PYTHIA.  For the polarization 
observables, the helicity asymmetry
distributions from Gehrmann and Stirling \cite{GS96} are 
evaluated at the Bjorken $x$ values ($x_{1(2)}$) 
given by PYTHIA, after being evolved \cite{HKM98} to 
$Q^2=p_{T,\gamma}^2 /2$.  
Gehrmann and Stirling provide three different sets of distributions,
referred to here as GS-A,B and C.  The sets differ mostly for $\Delta
G(x)$ (Fig.\ 2).
Armed with these variables, 
the proton spin correlation for a given $pp$ scattering 
event is calculated as:
$$A_{LL}={\Delta f_a(x_1,Q^2) \over f_a(x_1,Q^2)} 
{\Delta f_b(x_2,Q^2) \over f_b(x_2,Q^2)}
{\hat a}_{LL}^{proc}({\hat s},{\hat t},{\hat u}).  \eqno(5)$$
In this equation, $\Delta f/f$ is the polarization
of the parton (either a gluon or a quark (antiquark) of a given 
flavor) within one of the interacting protons 
(either beam $a$ or $b$), assuming the beams are 
100\% polarized.  The generated event may in principle arise from
colliding protons with either equal (+) or opposite (-) helicities,
but with different probabilities, proportional to
$$\mu_\pm=[1\pm P_{b_1}P_{b_2}A_{LL}]\sigma_{eff}. \eqno(6)$$
In the simulations, the initial helicity state for a given event is
chosen randomly in accordance with the above relative probabilities.
In Eqn. 6 the beam polarizations, $P_{b_{1(2)}}$, 
are each taken to be equal to 0.7, as expected for 
the RHIC-spin program.  Alternate selection between 
the two polarization states (representing for 
two-spin observables either equal or opposite helicity states
for the colliding protons) is 
continued until a non-zero value is drawn, 
corresponding to the occurrence of a collision.

A key assumption in this method of computing polarization observables,
is that initial-state parton showers are spin independent.  This
assumption has been checked in simulations (SPHINX \cite{SPHINX}) 
that separately consider parton showers 
independently for each helicity projection of the initial-state
partons before the hard scattering.  The good agreement \cite{Ma99} 
between SPHINX and the method outlined above for the polarization
observables suggests minimal spin-dependent effects from the
initial-state parton showers.  Further support for this conclusion
comes from next-to-leading order QCD calculations \cite{GV94}.
Higher-order processes are found to minimally influence the
polarization observables computed in leading order.  One precaution is
that, to date, the necessary resummation, required to account for
multiple soft gluon emission, has not yet been carried out for photon
production processes.

\begin{figure}
\epsfxsize=11.5cm \epsfbox{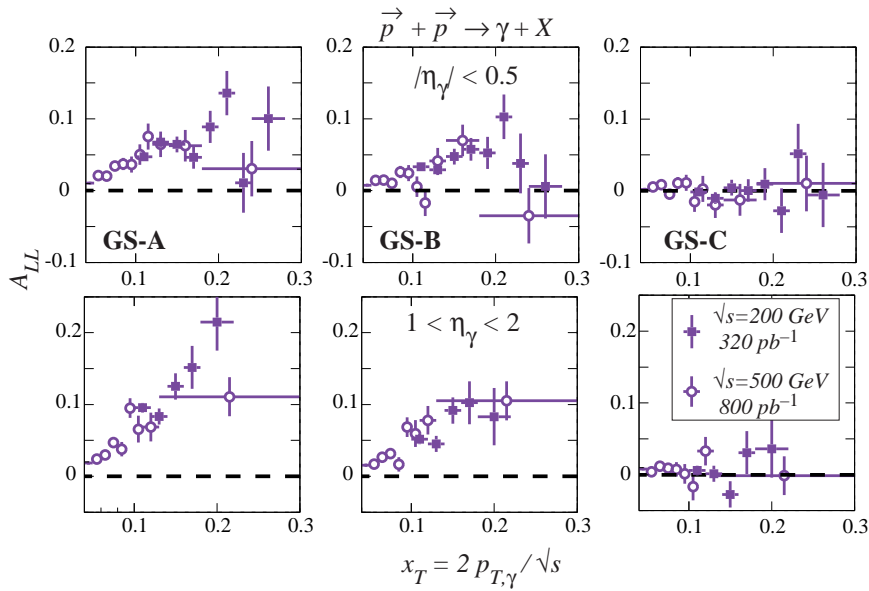}
\caption{Simulated values for $A_{LL}$ for inclusive photon production
at RHIC energies.  The top row shows the spin correlations for
midrapidity photons that can be detected at both PHENIX and STAR.
The bottom row shows expected values at the forward angles probed by
the STAR endcap electromagnetic calorimeter (EEMC).  Not evident in
this figure, is that for a given $x_T$, photons detected in the EEMC
correspond to smaller-$x$ gluons than those detected at midrapidity.}
\end{figure}

Given this methodology, polarization observables 
have been calculated for a variety of processes.  
The acceptance and finite 
resolution of the STAR barrel and 
endcap calorimeters (EMC) for photons
is accounted for in these simulations.
Isolation criteria \cite{UA293} are applied to the simulations,
anticipating their need in real experiments to reduce backgrounds
(Sect. 5.5).
Some of the calculations presented below rely on hadronic jet
reconstruction.  For those calculations, the response of the STAR TPC
and the EMC are approximately accounted for.  
Standard jet reconstruction algorithms \cite{SN196} are applied, 
based on the charged particle tracking from the STAR TPC
and the electromagnetic energy detection from the 
EMC, and assuming that long-lived neutral 
hadrons ($n,\overline{n},K_L^0$) are not detected.  
To avoid edge effects, the reconstructed jet is 
required to be within a full cone radius 
(taken to be of size, $R$ = 0.7 radians) from the edge of the STAR 
detector limits.  

The resulting asymmetries for inclusive direct photon 
production, under the conditions described above, are
shown in Fig. 4 as a function of the photon transverse momentum,
scaled by $\sqrt{s}$.  Several observations can be made:

\begin{itemize}
\item  For the GS-A,B gluon helicity asymmetry distributions, the $A_{LL}$
values are expected to be `large' ($>0.05$) for large $x_T$,
corresponding to values of $x_g$ where the gluon polarization is
large.  The small $A_{LL}$ values for GS-C reflect the small gluon
polarization at all $x_g$, arising because 
the peak of $x\Delta G(x)$ (Fig. 2) occurs at an
$x_g$ value where the unpolarized distribution is already quite large.

\item The qualitative features of direct photon production (discussed
in Sect. 5.1) are evident by comparing $A_{LL}$ for midrapidity and
forward-angle photons.  The latter, in general, results in larger
magnitude asymmetries; and hence, a greater sensitivity to $\Delta G$.

\item It is necessary to detect very hard photons ($p_{T,\gamma}>30$
GeV/c) to enable measurements of $A_{LL}$ at the same $x_T$ for two
values of $\sqrt{s}$.  Higher energy photon detection is more
difficult, but is important to accomplish.  Sampling the same $x_T$ at
different $\sqrt{s}$ is an important check of the theory.  Direct
photon production cross sections have been shown to deviate from
next-to-leading order QCD calculations in an $x_T$-dependent manner
\cite{CTEQ95}.  One speculation for this is the need for larger $k_T$
smearing than is present in the theory.

\end{itemize}

\subsection{Determining the initial-state partonic kinematics}
Direct photon production, where only the 
photon is detected (inclusive detection), provides only 
crude determination of the initial-state 
partonic kinematics.  It is generally assumed for photons 
detected at mid-rapidity ($\eta \approx 0$), 
that the kinematic quantity $x_T = 2p_{T,\gamma}/\sqrt{s}$ 
is approximately equal to the Bjorken $x$ values of the initial-state 
colliding partons.  This assumption is only approximately valid for
the {\it average values} of $x_g$ and $x_q$.  In fact, due to the lack
of any constraint on the direction of the recoiling jet, $x_g$ and
$x_q$ are distributed along an approximately hyperbolic locus at fixed
$x_T$ and $\eta_\gamma$.  This locus is slightly broadened by $k_T$
smearing effects.

These kinematic ambiguities can be vastly reduced by detecting the
away-side jet in coincidence with the photon.  With the reasonable assumption 
of collinearity of the colliding partons in the 
initial state, it is easy to show that the 
initial-state momentum fractions can be determined for each 
event by measuring the energy ($p_{T,\gamma}$)
and direction of the photon ($\eta_\gamma$), 
and only the direction of the away-side hadronic 
jet ($\eta_{jet}$).  With this information, 
conservation of energy and momentum at the partonic level imply
$$x_1={x_T \over 2}(e^{-\eta_{\gamma}}+e^{-\eta_{jet}})  {\rm ~~and~~}
x_2={x_T \over 2}(e^{+\eta_{\gamma}}+ e^{+\eta_{jet}}). \eqno(7)$$
Given the measured quark and gluon helicity-independent 
probability distributions, the reasonable assumption is made
that the initial-state quark had momentum 
fraction $x_q^{recon}={\rm max}[x_1,x_2]$ and the 
gluon, $x_g^{recon}={\rm min}[x_1,x_2]$.  Similarly, it is
straightforward to express the partonic CM scattering angle in terms
of $\eta_\gamma$ and $\eta_{jet}$.
A key assumption, 
that is examined more carefully below, is 
that the initial-state partons are collinear.  A valid 
question is, to what extent do transverse 
momentum components in the initial state ($k_T$) 
interfere with the event-by-event determination of 
the kinematics?

\begin{figure}
\epsfxsize=11.5cm \epsfbox{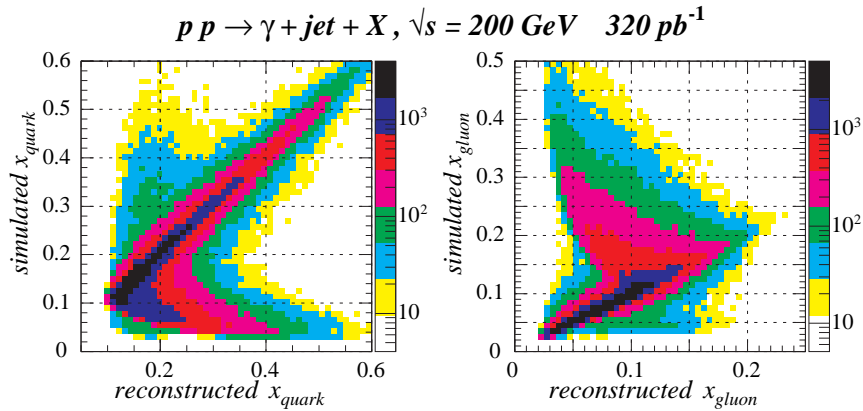}
\caption{Correlation between reconstructed and simulated values for
the quark and gluon momentum fractions for the 
$p+p\rightarrow \gamma+jet +X$ reaction.  The simulations include
effects of $k_T$ smearing.  The reconstruction assumes collinear
collisions.  Only events with an isolated photon in the STAR acceptance
($-1\leq\eta\leq 2$) having $10 \leq p_T \leq 20$ GeV/c are
accepted.  The coincident jet must be within the fiducial volume of
STAR ($-0.3 \leq \eta_{jet} \leq 1.3$).  A further restriction that
the reconstructed $x_{quark} \geq 0.2$ is also imposed.}
\end{figure}

To address this question, the reconstruction 
algorithm was applied to PYTHIA simulations of 
direct photon processes.  The initial-state 
partons had transverse momentum components, as 
generated by the parton shower model.  Events 
having at least one jet  and a coincident photon 
within the STAR fiducial volume were reconstructed.  
The UA2 isolation condition \cite{UA293} was applied to 
the photon.  A condition that the reconstructed 
$x_q$ was greater than 0.2 was also imposed.  As 
is evident in Fig. 5, most of the events have 
the initial-state momentum fractions properly 
reconstructed.  The majority of the kinematic reconstruction errors
occur when $x_q < x_g$.

\subsection{Direct reconstruction of $\Delta G(x)$}
Armed with the knowledge that the initial-state 
partonic kinematics can be accurately reconstructed, 
it is possible to consider {\it directly reconstructing 
$\Delta G(x)$ from the measured longitudinal spin 
correlation}.  It is possible that such a direct reconstruction will
provide the best framework for deducing $\Delta G$ from photon
production measurements.
As is already evident from Eqn. 5, 
if only quark-gluon Compton scattering contributes 
to the photon yield, then
$$A_{LL}={\Delta G(x_g,Q^2) \over G(x_g,Q^2)}
 A_1^p(x_q,Q^2) {\hat a}_{LL}^{Compton}(\theta^*).  \eqno(8)$$
This equation results from Eqn. 5, when averaging 
over an event ensemble, since the $u$ and $d$ 
quark contributions to gluon Compton scattering are weighted by their squared 
electric charge and the probability to find them 
inside the proton.  This is identical to the weighting that enters in
polarized deep-inelastic scattering (PDIS).  The quantity, $A_1^p$, in Eqn. 8
is then precisely the asymmetry measured in PDIS \cite{Pe99}.

\begin{figure}
\epsfxsize=11.5cm \epsfbox{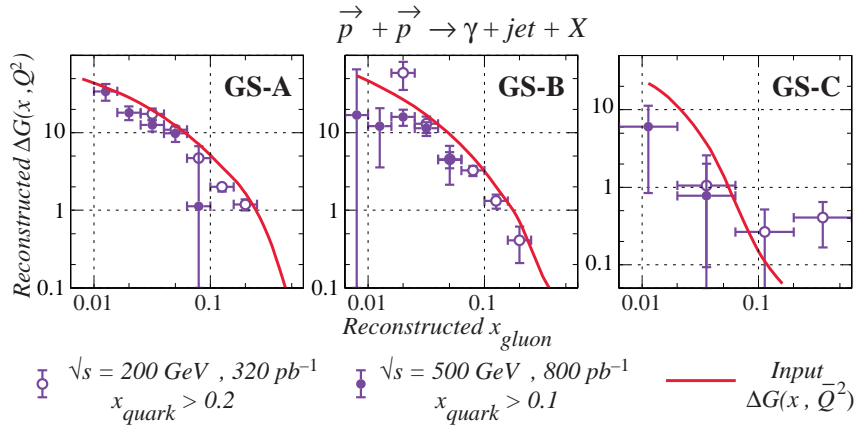}
\caption{Results for the direct reconstruction of $\Delta G(x,Q^2)$,
as applied to the simulated values of $A_{LL}$ for direct photon
processes.  Three different inputs for $\Delta G(x)$ are shown.}
\end{figure}

Since the kinematics can be determined for 
individual events, it is possible to invert Eqn. 8 to 
express the reconstructed gluon helicity asymmetry distribution
in terms of {\it known} 
quantities:  the unpolarized gluon PDF, the 
quark polarization and the pQCD expression for the 
gluon Compton scattering partonic spin 
correlation; and the measured spin-dependent scattering yields
($N_{++(+-)}$):
$$\Delta G_{recon}(x_g)= 
{[N_{++}(x_g) - N_{+-}(x_g)] \over P_{b_1} P_{b_2} 
\sum_{i=1}^{N_{++} + N_{+-}}[{A_{1_p}^{{\rm DIS}}(x_{q_i},Q_i^2) 
{\hat a}_{LL}^{{\rm Comp.}}(\theta_i^*)/G(x_{g_i},Q_i^2)}]}. \eqno(9)$$
This inversion, as applied to simulations of 
$A_{LL}$ for direct photon processes is shown in 
Fig. 6.  The key assumptions implicit to this 
direct reconstruction, and their influence on $\Delta G_{recon}(x)$, are:

\begin{itemize}
\item {\it Only the gluon Compton process 
contributes to the $\gamma$+{\rm jet} yield; the other 
processes that contribute to the simulated yields are assumed to be
absent in the reconstruction.}
The small contribution from $q\overline{q}\rightarrow \gamma g$ 
is negative, and is proportional to the product of the quark and 
antiquark polarizations at the asymmetric Bjorken $x$ 
values probed in the experiment.  This 
contribution results in $\Delta G_{recon}(x)$ 
underestimating the input value of $\Delta G(x)$.

\item {\it The initial-state partonic kinematics are assumed to be
perfectly reconstructed}.  For some events, especially the small
number with $x_g > x_q$, there are reconstruction errors.  Typically,
$A_{LL}$ for the $x_g$ value reconstructed for these events is 
smaller than for those events that don't have kinematic reconstruction
errors.  The end result is that $\Delta G_{recon}(x)$ underestimates
the input value of $\Delta G(x)$.

\item {\it The partonic collisions are assumed 
to be collinear, so ${\rm k_T}$ smearing effects are 
assumed to be absent}.  These effects introduce non-zero transverse
momentum in the initial state, contrary to the assumption of collinear
collisions.  The end result is to make correlated errors in the
reconstruction of $x_q$ and $x_g$.

\end{itemize}

Corrections to $\Delta G_{recon}(x)$ for these three effects can be 
made based on simulations.  The first correction requires no knowledge
of $\Delta G(x)$, whereas the latter two would need to be made in an
iterative fashion, since knowledge of $\Delta G(x)$ is required.

In addition to these three effects, present in the results shown in
Fig. 6, other effects can influence $\Delta G_{recon}(x)$ 
that will be determined
from ${\vec p}+{\vec p}\rightarrow \gamma+jet+X$ data.  In particular,
the $\gamma \gamma$ decay of a high-energy $\pi^0(\eta)$ meson can 
mimic the detector response of a single high-energy photon.
Simulations \cite{CDR99} have shown that this background results in a
smaller magnitude $A_{LL}$ compared to direct photon production, and
if not properly corrected, would result in an underestimate of
the true $\Delta G(x)$ via the direct reconstruction method.  Another
problem facing the RHIC-spin experiments, is the possibility that some
fraction of the photons produced in ${\vec p}+{\vec p}$ collisions
arise from the `fragmentation' of recoiling final state partons.  In
particular, high-$p_T$ photons can be produced in the hard
bremsstrahlung of a charged parton, produced by $qg$ or $gg$
scattering processes that have significantly larger cross section than
the direct photon processes.  Simulations \cite{Bl99} have shown
that these other processes will dilute the direct photon $A_{LL}$ by a
small amount.  Again, they tend to underestimate the true $\Delta
G(x)$ via the direct reconstruction method.

\subsection{Backgrounds and their suppression}
To employ direct photon production as a means 
of determining the polarization of the proton's 
glue, a detector must be capable of selecting the 
very small fraction of events that have a single 
high energy photon.  The most pernicious physics 
background arises from di-jet events (by far, the 
largest fraction of the total reaction cross section), 
where one of the jets has an energetic neutral 
meson ($\pi^0$ or $\eta^0$, collectively referred 
to as $M^0$) that can decay into a 
pair of photons.  Kinematically, the most probable 
opening angle between the photons produced 
by $M^0$ decay is $\phi_{\gamma \gamma}^{\rm min} 
= 2sin^{-1}(m_{M^0}c^2/E_{M^0})$.
For a 30 GeV $\pi^0$ this angle is 9 mr.  Such a 
small opening angle makes it very difficult for 
any detector to distinguish between di-photons from 
$M^0$ decay and a single direct photon.

The relative probability for producing a neutral meson 
versus a direct photon is shown in Fig. 7a.  
Due to the larger number of processes that can 
produce jets versus single photons, and the 
fact that the strong interaction coupling constant 
($\alpha_S$) is involved rather than the 
electromagnetic coupling constant ($\alpha$),~$M^0$ 
production is nearly an order of magnitude 
larger than $\gamma$ production.  This result from 
PYTHIA has been shown \cite{Bl99} to be roughly in accord 
with existing measurements \cite{SN77}.

The influence of the $M^0$ background on the 
direct photon measurements at RHIC is different 
for the PHENIX and STAR detectors.  The former has a 
smaller solid angle, but substantially finer 
granularity, EMC than the latter.  This feature allows 
PHENIX to better reconstruct $M^0$ from their daughter 
photons.  The uniform distribution of the decay photon 
energies from $M^0$ decay ultimately sets a limit to the range of
decay phase space where 
this background suppression technique is effective.  The STAR 
detector will distinguish background from signal using two primary 
techniques, described below.

\begin{wrapfigure}{r}{6cm}
\epsfig{figure=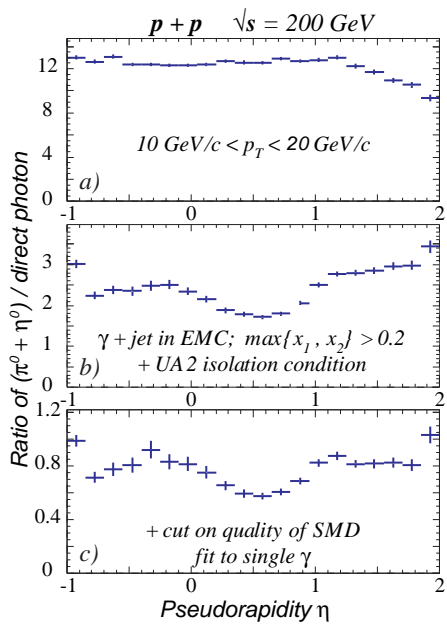,width=6cm}
\caption{Ratio of cross sections for $\pi^0(\eta)$ and direct photon
production under increasingly stringent conditions:  a) including only
a $p_T$ cut; b) cuts on kinematic quantities and an isolation cut; c)
cut on information from SMD.}
\end{wrapfigure}

The first of the background suppression methods 
for STAR relies on a `shower maximum detector' (SMD)\cite{CDR99}.  
This is a fine granularity detector placed within 
the depth of the calorimeter to measure 
the transverse profile of the electromagnetic (EM) 
showers produced by the incident particles.  
In general, an incident $M^0$ results in two 
separated clumps of energy deposition in the SMD, corresponding to the 
spatial separation of the EM showers produced by 
the closely spaced daughter photons.  An incident direct
photon typically produces only a single 
clump of energy deposition in the SMD \cite{CDR99}.  This 
distinction is muddied by the complexity of EM 
showers, resulting in false clumpiness in the SMD 
response for some fraction of the direct-photon-induced events.  
Despite this, the analysis of the SMD information 
will provide significant suppression of the $M^0$ background.

The second component of the background suppression arsenal 
of STAR is related to the nearly 4$\pi$ coverage 
the detector provides.  In general, high energy neutral 
mesons are leading particles of jets, and hence, are 
accompanied by additional hadrons.  An effective 
discrimination between $M^0$ and direct photon events is 
provided by examining the event topology.  The former class of 
events generally have additional hadrons within a 
relatively narrow cone around the $M^0$ whereas direct 
photon events have little accompanying energy within the 
isolation cone.  The $M^0$ events can be suppressed by 
using `isolation cuts' similar to those used in other collider 
detectors reporting direct photon cross sections.  

The influence of these background reduction cuts 
on the background:signal ratio is illustrated in 
Figs. 7b and 7c.  The end result is that the signal is 
expected to be approximately 2$\times$ larger
than the background.  The remaining background contributions 
can be eliminated by taking the difference 
between distributions enriched in direct photons and 
those enriched in $M^0$.  This subtraction 
process will increase the statistical errors beyond 
those shown in Fig. 6 by a factor varying
between 1.5 and 2 \cite{CDR99}.  The systematic error associated with 
these corrections can be estimated from 
{\it in situ} calibrations of the performance of the 
SMD, made possible by producing an energetic  
$\pi^0$ sample from reconstructed $\rho^\pm$ events.  
Other backgrounds will also be present in 
both the STAR and PHENIX experiments, but their effects 
on the determination of $\Delta G$
should be less important.  Efforts are underway to 
understand the limitations on extracting 
$\Delta G$ from direct photon measurements.

\subsection{Extrapolation Errors}
Unlike the goals of most experiments sensitive to 
gluon polarization that are either underway \cite{Br99} or are 
on the horizon \cite{BHK98}, the RHIC-spin program aims to 
determine the {\it integral contribution} gluons make 
to the proton's spin, as defined in Eqn. 1.  
To carry out this integral, the gluon helicity asymmetry 
distribution must be determined over a sufficiently 
broad range of $x_g$ to minimize the influence 
of extrapolation errors in carrying out the integral over all $x$.  
To illustrate how these 
errors influence the determination of $\Delta G$, a 
standard parameterization of the $x_g$ dependence,
$$x\Delta G(x) = \eta A x^a (1-x)^b [1 + \rho x^{1/2} + \gamma x],$$
$$ {\rm with~~} A^{-1}=\left( 1+{\gamma a 
\over a+b+1}\right){\Gamma(a) \Gamma(b+1) 
\over \Gamma(a+b+1)}
+ \rho {\Gamma(a+{1 \over 2})\Gamma(b+1) \over \Gamma(a+b+1)},  \eqno(10)$$
$${\rm so~that~~}
\eta = \int_0^1 \Delta G(x) dx,$$
is fit to simulated data.  Similar to the method used to analyze
\begin{figure}
\epsfxsize=11.5cm \epsfbox{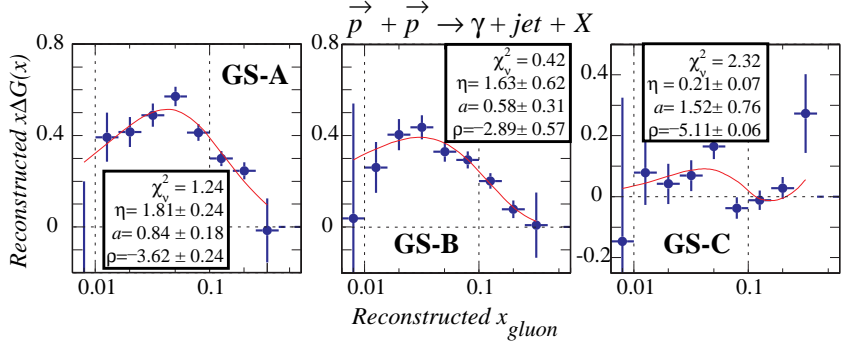}
\caption{Fits to the reconstructed $\Delta G(x)$, after correcting for
$q\overline{q}$ annihilation.  A standard parameterization (Eqn. 10)
is used to fit the data, with the $b$ and $\gamma$ parameters fixed.
Full data sets for $\sqrt{s}$=200(500) GeV are assumed, corresponding
to integrated luminosity of 320(800) pb$^{-1}$.}
\end{figure}
existing data on scaling violations in polarized deep inelastic
scattering \cite{GS96}, $b$ and $\gamma$ are held fixed at values
obtained by evolving \cite{HKM98} the $\Delta G(x)$ input to the
simulation to the $Q^2$ values relevant at RHIC.  
The fixed parameters used in the fits are consistent with positivity
constraints ($|\Delta G(x)| < G(x)$).  The parameters
$\eta, a,\rho$ are then adjusted to provide the best fit to $\Delta
G_{recon}(x)$, described in Sect. 5.4.  The results for the
simulations of the $\sqrt{s}$=200 and 500 GeV samples are combined,
after making additive corrections for $q\overline{q}$ annihilation
contributions to the simulated asymmetries.
No corrections have been made for kinematic reconstruction errors or
$k_T$ smearing.  As well, the values of $\Delta G_{recon}(x)$ have not
been evolved to a common $Q^2$.  These corrections are important, but
have not been made because they require knowledge of $\Delta G(x)$.
The fits are shown in Fig. 8.  It is found that:
\begin{itemize}
\item {\it an accurate determination of $\Delta G$ will require both
$\sqrt{s}$=200 and 500 GeV data samples to get to sufficiently small
$x_g$.}  Due to
strong correlations between $\eta$ and $a$, $\delta \eta$ grows
rapidly as the low-$x$ points are successively eliminated.
The error in the integral $\Delta G$ ($\delta \eta$), with both samples
included in the fit, is 0.24 when Gehrmann-Stirling
\cite{GS96} set A is input to the simulations.  The value for
$\delta \eta$ for set B is 0.62, and the relative error in $\eta$ is comparable
for set C.  For set A, other systematic errors are expected\cite{CDR99} to 
increase the error in the integral $\Delta G$ to 0.5, 
still sufficient to accurately establish 
$\Delta \Sigma$ from polarized deep inelastic scattering.  

\item {\it the large-$x$ behavior of $\Delta G(x)$ must be constrained to
determine $\Delta G$ at RHIC}.  Different values for the fixed
parameters ($b,\gamma$) result in different values for $\eta$.

\item {\it fitting $\Delta G_{recon}(x)$, including only the
corrections for $q\overline{q}$ annihilation, yields a value for the
fitted $\eta$ that is too small compared to the input $\Delta G$.}
Evolving all of the $\Delta G_{recon}(x)$
points to a common $Q^2$ and correcting for the kinematic
reconstruction errors results in a fitted $\eta$ that is closer to the input
$\Delta G$, but is still too small.  Both of these corrections require
knowledge of $\Delta G(x)$, and hence will require an iterative
procedure for their application.  The largest remaining error comes
from $k_T$ smearing.  Repeating the analysis with simulations that
don't include initial-state parton showers results in a fitted $\eta$
in agreement with the input $\Delta G$.
\end{itemize}

The end result is, that after accounting for the most significant
sources of systematic error \cite{CDR99}, we expect that the fraction
of the proton's spin carried by gluons can be determined to an
accuracy of approximately 0.5, primarily based on the STAR measurements of
$\vec{p}+\vec{p}\rightarrow \gamma+{\rm jet}+X$.  Data samples at both
$\sqrt{s}$=200 and 500 GeV are crucial so that the accuracy is not
limited by extrapolation errors.  The analysis of $\Delta
G_{recon}(x)$ presented here is intended to illustrate the sensitivity
of the STAR measurements to the integral $\Delta G$.  Clearly, the best
determination of $\Delta G$ will result from a global analysis of all
relevant data.  

\section{Summary}
The RHIC-spin program promises to provide exciting data to test
perturbative QCD, and to provide new insights into the
non-perturbative structure of the proton.  It is likely that one of
the most interesting results will be the determination of the fraction
of the proton's spin carried by gluons.  
I expect that inclusion of $\Delta G_{recon}(x)$, 
based on the measurement of $A_{LL}$ for 
the $\vec{p}+\vec{p} \rightarrow \gamma + jet + X$ reaction at
$\sqrt{s}$=200 and 500 GeV, 
in a global analysis will provide the best determination of the gluon 
contribution to the proton's spin. 

\section*{Acknowledgments}
I gratefully acknowledge the extensive discussions I've had with my
IUCF colleagues (W.W. Jacobs, J. Sowinski, E.J. Stephenson,
S.E. Vigdor and S.W. Wissink) while carrying out this work.  I also thank
my colleagues for their careful reading of, and comments on, this manuscript.

\end{document}